\documentclass{ifacconf}

\usepackage{graphicx}      
\usepackage{natbib}        

\makeatletter
\let\old@ssect\@ssect 
\makeatother

\usepackage[colorlinks,urlcolor=blue,linkcolor=blue,citecolor=blue]{hyperref}

\makeatletter
\def\@ssect#1#2#3#4#5#6{%
  \NR@gettitle{#6}
  \old@ssect{#1}{#2}{#3}{#4}{#5}{#6}
}
\makeatother

\usepackage{color,array}

\usepackage{amsfonts}
\usepackage{subfigure}
\graphicspath{{./IMG/}}
\usepackage{epstopdf}
\usepackage{comment}
\usepackage{amsmath}
\usepackage{etoolbox}
\usepackage{algorithm}

\let\theoremstyle\relax 
\usepackage{amsthm}

\AtBeginEnvironment{algorithm2e}{\algoequations}
\AtEndEnvironment{algorithm2e}{\restoreequations}
\newcounter{algosavedequation}
\newcommand{\algoequations}{%
  \setcounter{algosavedequation}{\value{equation}+1}%
  \setcounter{equation}{0}%
  \renewcommand{\theequation}{\arabic{algosavedequation}\alph{equation}}
}
\newcommand{\restoreequations}{%
  \setcounter{equation}{\value{algosavedequation}}%
}

\allowdisplaybreaks[3]

\newtheorem{assumption}{Assumption}
\newtheorem{problem}{Problem}

\newtheorem{remark}{Remark}
\newtheorem{theorem}{Theorem}

\newcommand{\mc}{\mathcal}

\newcommand{\B}{\mathbb{B}} 					

\newcommand{\real}{\mathbb{R}}

\newcommand{\realpos}{\mathbb{R}_{> 0}}
\newcommand{\realnneg}{\mathbb{R}_{\geq 0}}

\newcommand*{\QEDB}{\hfill\ensuremath{\square}}
\newcommand*{\QEDBL}{\hfill\ensuremath{\blacksquare}}
\newcommand\oprocendsymbol{\hbox{$\square$}}
\newcommand\oprocend{\relax\ifmmode\else\unskip\hfill%
\fi\oprocendsymbol}

\newcommand{\R}{\mathbb{R}} 
\newcommand{\proj}{\Pi_{\mathcal{U}_c}}

\newcommand{\map}[3]{#1: #2 \rightarrow #3}

\newcommand{\sbs}[2]{{#1}_{\textup{#2}}}

\newcommand{\U}{\mathcal{U}}
\newcommand{\W}{\mathcal{W}}

\newcommand{\X}{\mathcal{X}}

\begin{document}
\begin{frontmatter}

\title{Perception-Based Sampled-Data Optimization of Dynamical Systems\thanksref{footnoteinfo}}

\thanks[footnoteinfo]{This work was supported in part by the National Science Foundation through the award CMMI 2044946. Corresponding author: Liliaokeawawa Cothren: Liliaokeawawa.Cothren@colorado.edu}

\author[First]{Liliaokeawawa Cothren} 
\author[Second]{Gianluca Bianchin} 
\author[Third]{Sarah Dean}
\author[First]{Emiliano Dall'Anese}

\address[First]{Department of ECEE,
        University of Colorado Boulder}
\address[Second]{ICTEAM Institute, Universit\'{e} Catholique de Louvain }
\address[Third]{Department of Computer Science,
        Cornell University}

\begin{abstract} 
Motivated by perception-based and sensing-based control problems in autonomous systems, this paper addresses the problem of developing feedback controllers to regulate the inputs and the states of a dynamical system to optimal solutions of an optimization problem when 
one has no access to exact measurements of the system states. In particular, we consider the case where the states need to be estimated from high-dimensional sensory data received only at some  time instants. We develop a sampled-data feedback controller that is based on adaptations of a projected gradient descent method
and includes neural networks as integral components to estimate the state of the system from 
perceptual information. We derive sufficient conditions to guarantee (local) input-to-state stability of the control loop. Moreover, we show that the interconnected system tracks the solution trajectory of the underlying optimization problem up to an error that depends on the approximation errors of the neural network and on the time-variability of the optimization problem; the latter originates from time-varying safety and performance objectives,  input constraints, and unknown disturbances. 
\end{abstract}

\end{frontmatter}

\section{Introduction}\label{S:introduction}


A major challenge in controlling complex autonomous systems consists of incorporating rich  data from perceptual and sensing sensor data.  The performance of feedback control systems critically relies on the information extracted from perceptual sensing, which may require the processing of high-dimensional data only available at given spatio-temporal granularities; see, e.g.~\cite{dean2021certainty,al2020accuracy,xu2021learned,dawson2022learning}. This paper investigates how to integrate perceptual information into controllers inspired by optimization algorithms where the goal is to steer a dynamic system toward solutions of an optimization problem with costs associated with the system's inputs and states. For example, in autonomous driving, the optimization problem may formalize way-point tracking and obstacle avoidance, whose information is provided by images from a dashboard camera. Other examples include robotics and power systems (the latter leveraging pseudo-measurements). 

The line of research on feedback optimization goes back to earlier concepts of KKT-type controllers in~\cite{Jokic2009controller,brunner2012feedback, Hirata}, and it was recently expanded to include new classes of controllers inspired by first-order optimization methods in~\cite{MC-ED-AB:20,zheng2019implicit,hauswirth2020timescale,lawrence2020linear,bianchin2021time,bianchin2022online,SampledDataOnlineFeedbackEquilibriumSeeking_2021,agarwal2022game,simpson2021low}; also see references therein. In this paper, we provide contributions relative to existing works by considering a setup where the cost of the optimization problem evolves over time (for instance, for way-point tracking and to avoid moving obstacles) and the state of the system cannot be directly measured. The latter is a distinctive feature of this work: we address the case in which optimization-based controllers must leverage perceptual information available at given temporal granularities and learning mechanisms to estimate the system state.

We develop a sampled-data feedback controller that is based on an adaptation of a projected gradient descent method. Based on the specified time-varying costs, the gradient-based controller generates inputs for the system which are then passed through a zero-order hold. Importantly, the controller leverages a trained neural network that maps perceptual information into estimates of the state of the system. We derive sufficient conditions to guarantee (local) input-to-state stability (ISS) of the control loop. In particular, we show that the interconnected system tracks the optimal solution trajectory of the optimization problem up to an error that depends on the approximation errors of the neural network and the time-variability of the cost and unknown disturbance. The ISS bounds are derived by leveraging the fundamental results of~\cite{jiang2001input} and~\cite{nevsic1999formulas}.  

We note that a similar perception-based regulation problem was considered in our 
prior work in~\cite{cothren2022online}; however, in~\cite{cothren2022online} 
controllers operate at continuous time and thus do not account for the sample-data 
nature of the feedback information. A sampled-data controller was developed in~\cite{SampledDataOnlineFeedbackEquilibriumSeeking_2021}; with respect to~\cite{SampledDataOnlineFeedbackEquilibriumSeeking_2021}, we consider cases 
where the optimization problem is time-varying and the state is estimated via perception maps. 

We test our controller on an autonomous driving application where vehicles are modeled via unicycle dynamics; the controller acquires the position of the vehicle from a neural network that estimates positions from images.  


\section{Problem Formulation}\label{S:problem-formulation}
In the following, we formalize our research problem and discuss the necessary 
%
%
assumptions.\footnote{\emph{Notation}. We denote by \( \mathbb{N}, \mathbb{Z}_{+}, \R, \realpos, \text{ and } \realnneg \) the set of natural numbers, positive integers, real numbers, positive real numbers, and non-negative real numbers. For vectors \( x \in \R^n \) and \( u \in \R^m \),
\( \| x \| \) is the  Euclidean norm of $x$, \( \| x \|_\infty \) is the 
supremum norm, and \( (x,u) \in \R^{n + m}\) is their vector concatenation. $x^\top$ denotes transposition, and $x_i$ denotes the $i$-th element of $x$. For a matrix $A \in \R^{n \times m}$, $\|A\|$ is the induced $2$-norm and $\|A\|_\infty$ is the supremum norm. 
The set $\mathcal{B}_n(r) := \{ z \in \R^{n} : \|z\| < r \}$ is the open ball in $\R^{n}$ with radius $r > 0$; $\mathcal{B}_n[r] := \{ z \in \R^{n} : \|z\| \leq r \}$ is the closed ball. Given two sets $\mathcal{X} \subset  \mathbb{R}^n$ and  $\mathcal{Y} \subset  \mathbb{R}^m$, $\mathcal{X} \times \mathcal{Y}$ is their Cartesian product; $\mathcal{X} + \mathcal{B}_n(r)$ is the open set defined as $\mathcal{X} + \mathcal{B}_n(r) = \{x + y: x \in \mathcal{X}, y \in \mathcal{B}_n(r)\}$. 
$\Pi_{\mathcal{U}}$ is the Euclidean projection of $z \in \R^n$ onto $\mathcal{U}\subset \R^n$; or, $\Pi_{\mathcal{U}}(z) := \operatorname{arg}\min_{u \in \mathcal{U}} \|u - z\|^2$. A function $\map{\gamma}{\realnneg}{\realnneg}$ is of class $\mc K$ if 
it is continuous, $\gamma(0) = 0$, and strictly increasing; it is of 
class $\mc K_\infty$ if it is additionally unbounded. A function $\beta: \realnneg \times \realnneg \to \realnneg$ is of class $\mathcal{KL}$ if for each fixed $t$ the function $\beta(r,t)$ is of class $\mc K$, and if for each fixed $r$ the function $\beta(r,t)$ is decreasing with respect to $t$ and is s.t. $\beta (r,t)\rightarrow 0$ for $t\rightarrow \infty $.}
We consider systems that can be modeled using dynamics of the form: 
\begin{align}\label{eq:general-plant}
    \dot x = f(x,u,w), \,\, x(0) = x_0,
\end{align}
where $x: \R_{\geq 0} \to \X \subseteq \R^{n_x}$ is the state,
$u: \R_{\geq 0} \to \U \subseteq \R^{n_u}$ is the control input, 
$w: \R_{\geq 0} \to \W \subseteq \R^{n_w}$ is a time-varying unknown exogenous disturbance, and where the vector field $f: \X \times \U \times \W \to \R^{n_x}$ is continuously differentiable on the open and connected domain $\X \times \U \times \W$ and is Lipschitz-continuous in its arguments with constants $L_x, L_u$, and $L_w$ respectively. In this paper, motivated by practical hardware and operational requirements,  we restrict our attention to cases where  $u \in \sbs{\mc U}{c}$ at all times, where $\sbs{\mc U}{c} \subset \mc U$ is  compact. We impose the following additional assumptions on the system \eqref{eq:general-plant}. 

\vspace{.1cm}

\begin{assumption}\label{as:steady-state-map}
There exists a unique continuously differentiable map $h: \U \times \W \to \X$ such that, for any (constant)  $\bar u \in \U$ and $\bar w \in \W$, \( f\left( h(\bar u, \bar w), \bar u, \bar w \right) = 0.\)
Moreover, $h(u,w)$ admits the decomposition $h(u,w) = h_u(u) + h_w(w),$ where $h_u$ and $h_w$ are Lipschitz continuous with constants $\ell_{h_u}$ and  $\ell_{h_w}$, respectively. \hfill \QEDB
\end{assumption}

\vspace{.1cm}

\begin{assumption}\label{as:restrict-control-inputs}
For all $t \in \realnneg$, $w(t) \in \sbs{\mathcal{W}}{c}$, where 
$\sbs{\mathcal{W}}{c} \subset  \mathcal{W}$ is compact. Moreover,
\( t \mapsto  w(t)\) is continuous. 
\QEDB
\end{assumption} 

Assumption \ref{as:steady-state-map} guarantees that, for constant inputs $\bar u$ and $\bar w$, system \eqref{eq:general-plant} admits the unique equilibrium point $\bar x := h(\bar u, \bar w)$. 
Note that when $\nabla_x f(x, \bar u, \bar w)$ is invertible for any $\bar u$ and $\bar w$, then the existence of $h(\bar u, \bar w)$ is always guaranteed. Furthermore, by the implicit function theorem, $h(\bar u, \bar w)$ is differentiable since $f(x,\bar u, \bar w)$ is differentiable. 
We also note that the equilibrium set $\X_{eq} := \{ h(\bar u, \bar w) : \bar u \in \U_c, \,\, \bar w \in \mc W_c \}$ is compact; this is due to $\U_c \times \W_c$ being compact, $h(u, w)$ being continuously differentiable, and the result \cite[Thm. 4.14]{rudin1976principles}. For any $u \in \U_c$, we have that $\| \nabla_u h(u,\bar w)\| \leq \ell_{h_u}$ since $\U_c$ is compact \cite{rudin1976principles}. \hfill

Next, as customary in the context of feedback optimization (see, e.g.,~\cite{MC-ED-AB:20,hauswirth2020timescale}, we require a stability condition on the system to control. To this end, let  $\mathcal{X}_r = \mathcal{X}_{eq} + \mathcal{B}_r(r) \subseteq \X$, $r > 0$, be a set
for which the following assumption holds. 

\vspace{.1cm}

\begin{assumption}\label{as:class-K-straight-assumption}
There exists a continuously differentiable function $V: \X_r \times \U \times \W \to \R$ with constants $d_1, d_2, d_3 > 0$ and a $\mathcal{K}$-function $\sigma_w$ such that: 
\begin{enumerate}
    \item For all $x\in \X$, $u \in \U$, and $w \in \W$,
    \[ \quad\quad\quad  d_1 \|\tilde x\|^2  \leq V(\tilde x,u,w) \leq d_2 \|\tilde x\|^2,\]
    where $\tilde x := x - h(u,w)$; 
    \item For any constant $u \in \U$,
    \[ \dot V(x(t), u, w(t)) \leq -d_3 V(x(t), u, w(t)) + \sigma_w(\|\dot w(t)\|). \QEDB \]
\end{enumerate}
\end{assumption}

\vspace{.1cm}

From \cite{Khalil:1173048},
Assumption \ref{as:class-K-straight-assumption} implies that there exists constants $k, a, \gamma > 0$ such that the following holds: 
\begin{align*}
    \|\tilde x(t)\| \leq k \|\tilde x(0)\| e^{-at} + \gamma \sigma_w \left( \sup_{0\leq \tau\leq t} (\|\dot w(\tau)\|) \right),
\end{align*} 
for some constant $\gamma > 0$, and for $x(0) \in \X_0 := \X_{eq} + \B_n(r_0), \,\, r_0 < (r-\text{diam}(\sbs{\mathcal{X}}{eq} ))/k - \bar \gamma$, $\bar \gamma := \gamma \sigma_w \left( \sup_{t \geq 0} \|\dot w(t)\|)\right)$.


We point out that when the (physical) system does  not satisfy 
Assumption \ref{as:class-K-straight-assumption}, then~\eqref{eq:general-plant} 
models the pre-stabilized physical system.

\vspace{-.1cm}

\subsection{Generative and Perception Maps}\label{SS:perceptionMaps}

\vspace{-.25cm}

In this paper, we assume that the state $x$ of~\eqref{eq:general-plant} is not directly measurable. Instead, one has access to nonlinear and possibly high-dimensional observations of the
state  $\zeta = q(x)$, where $q: \mathcal{X} \rightarrow \real^{n_\zeta}$ is an \emph{unknown} generative map. This setup emerges when information about the state is acquired through  perceptual information from sensing and estimation mechanisms. For example,  in applications in autonomous driving, vehicle states are often reconstructed from images generated by cameras. See, for example, the models in~\cite{dean2021certainty,al2020accuracy,murillo2022data}; also see the closely related observer design problems in~\cite{Marchi2022,chou2022safe}.

Regarding the unknown map $x \mapsto q(x)$, we make the following assumption (see also~\cite{dean2021certainty}). 

\vspace{.1cm}

\begin{assumption}\label{as:observable-Map}
The map $q: \mathcal{X} \to \R^{n_\zeta}$ is such that the image of $q(\mathcal{X}')$ is compact for any compact set $\mathcal{X}' \subset \mathcal{X}$.  
Further, there exists a map $p: \R^{n_\zeta} \to \R^{n_x}$ such that $p(\zeta) = p(q(x)) = x + \varepsilon(x)$, where 
where $\varepsilon(x)$ is bounded as $\|\varepsilon(x)\| \leq \bar \varepsilon$ for any $x \in \mathcal{X}$, for a given finite $\bar \varepsilon \geq 0$.
\QEDB \end{assumption}

\vspace{.1cm}


The function $p(\zeta)$ 
in Assumption \ref{as:observable-Map} is referred to as the perception map; for a  given  observation $\zeta$, it yields a possibly noisy estimate of the state, up to a bounded error $\varepsilon(x)$. In this paper, we will leverage supervised learning methods to estimate the perception map $p(\zeta)$ from data. 

\vspace{-.1cm}

\subsection{Regulation to Solutions of an Optimization Problem}\label{SS:target-control-problem}

\vspace{-.1cm}

We focus on regulating the system \eqref{eq:general-plant} to the solution of the following time-varying optimization problem:
\begin{subequations}\label{opt:objectiveproblem}
\begin{align}
\label{opt:objectiveproblem-a}
(u^*(t), x^*(t)) \in  \arg
\underset{\bar u \in \mathcal{U}_c}{\min}  ~~ & 
\phi (\bar u, t) + \psi (\bar x, t)\\
\label{opt:objectiveproblem-b}
\text{s.t.} ~~~ & \bar x = h(\bar u, w(t)),
\end{align}
\end{subequations}
where $u \to \phi(u,t)$ and $x \to \psi(x,t)$ are functions that
describe costs associated with the system's inputs and states, respectively. 
We remark that~\eqref{opt:objectiveproblem} is a time-varying optimization 
problem for two reasons: (i) the cost functions are time-varying, which allow us 
to account for performance and safety objectives that evolve over time, and 
(ii) the constraint is time-varying since the system's equilibrium point is 
parametrized by the time-varying signal $w(t)$. Accordingly,~\eqref{opt:objectiveproblem} defines optimal trajectories $t \mapsto (u^*(t), x^*(t))$ for the system \eqref{eq:general-plant}. Note that, since $h(u,w)$ is unique for any fixed $u$ and $w$ (see Assumption~\ref{as:steady-state-map}), the optimization problem \eqref{opt:objectiveproblem} can be rewritten as: 

\vspace{-.35cm}
\begin{align}\label{opt:optimization-problem-ss}
    u^*(t) \in \arg\underset{\bar u \in \mathcal{U}_c}{\min} \, \phi(\bar u, t) + \psi\left( h(\bar u, w(t)), t \right).
\end{align}

\vspace{-.25cm}

Given the problem~\eqref{opt:optimization-problem-ss}  we formalize our control problem.

\vspace{.1cm}

\begin{problem}[\textbf{\textit{Online optimization with state perception}}]
Design an output feedback controller so that the inputs and states 
of~\eqref{eq:general-plant} track the time-varying solution $(u^*(t),x^*(t))$ of~\eqref{opt:optimization-problem-ss} when $w(t)$ in unknown and $x$ is not 
measurable; instead, we have access only to state estimates 
$\hat{x} = \hat p(\zeta)$ at certain instants  $\mathbb{S} = \{ k \tau : k \in \mathbb{Z}_{+} \}$, $\tau > 0,$ where $\zeta= q(x)$ and $\hat p(\cdot)$ is an estimate of the map $p(\cdot)$. 
\QEDB
\end{problem}


\vspace{.1cm}

\begin{remark}\textbf{(Implicit solution)}
\label{rem:time-variance-in-costs}
Since the problem~\eqref{opt:optimization-problem-ss} is parametrized by the unknown exogenous inputs $w(t)$, the solutions of~\eqref{opt:optimization-problem-ss} cannot be computed explicitly via standard numerical optimization methods. We seek feedback controllers that drive inputs and states of~\eqref{eq:general-plant} to solutions  of~\eqref{opt:optimization-problem-ss} by relying only on estimates of the state $\hat{x} = \hat p(\zeta)$, and without requiring sensing of the disturbance $w(t)$
. \hfill \QEDB
\end{remark}

\vspace{.1cm}

\begin{remark}\textbf{(Interpretation of the control problem)}
Recall that~\eqref{opt:optimization-problem-ss} is parametrized by a time-varying disturbance $w_t$ and has time-varying costs. 
Thus,~\eqref{opt:optimization-problem-ss} formalizes an 
\emph{equilibrium seeking problem}, where the objective is to select
 optimal input-state pairs $(u^*(t),x^*(t))$ that  minimizes 
the specified time-varying cost at each time $t$ (see, e.g.,~\cite{MC-ED-AB:20,bianchin2021time,SampledDataOnlineFeedbackEquilibriumSeeking_2021}). This is a high-level regulation problem that 
can be nested with a stabilizing controller.
\hfill \QEDB
\end{remark}

\vspace{.1cm}

We conclude this section with some relevant assumptions. 

\vspace{.1cm}

\begin{assumption}\label{as:costFunctionAssumptions}
The following hold: 

\ref{as:costFunctionAssumptions}(i) The function $u \mapsto \nabla \phi(u,t)$ is $\ell_u$-Lipschitz continuous for all $u \in \U$, $\ell_u \geq 0$, for all $t$.

\ref{as:costFunctionAssumptions}(ii) The function $x \mapsto \nabla \psi(x,t)$ is $\ell_x$-Lipschitz continuous for all $x \in \X$, $\ell_x \geq 0$, for all $t$.

\ref{as:costFunctionAssumptions}(iii) The function $u \mapsto \nabla \phi(u,t) + \partial_u h(u,w)^\top \nabla \psi(h(u,w),t)$ 
    is $\mu$-strongly convex with $\mu > 0$, for all $u \in \U$ and for all $t$, where $\partial_u h(u,w)$ is the Jacobian of $h(u,w)$ w.r.t. $u$. 
    
\ref{as:costFunctionAssumptions}(iv) The set $\U_c \subset \mathbb{R}^{n_u}$ is convex. \hfill \QEDB  
\end{assumption}

\vspace{.1cm}

Note that, from Assumption \ref{as:costFunctionAssumptions}, it follows that the composite cost $u \mapsto \nabla \psi(u) + \partial_uh(u,w)^\top \nabla \psi(h(u,w))$ is $\ell$-Lipschitz continuous with constant $\ell := \ell_u + \ell_{h_u}^2 \ell_x$. 


\section{Perception-Based System Regulation}\label{SS:optimal-regulation-with-perception-in-the-loop}
To address Problem~1, we consider the design of feedback controllers that are inspired by projected-gradient-type methods as in~\cite{bianchin2021time,cothren2022online}. However, to acknowledge the fact that the state estimates are available only at given time intervals (for example, images from a camera are captured at a given frequency), our controller design is based on a sampled data mechanism. The controller is equipped with a supervised learning method to estimate the perception map.

Towards this, let $\tau  > 0$ represent the period between two consecutive  
arrivals of perceptual data (for example, images) and let $k \in \mathbb{Z}_{+}$ be the sampling index, so that $\mathbb{S} = \{ k \tau : k \in \mathbb{Z}_{+} \}$ is the set of times where perceptual information arrives and control inputs are updated. Accordingly, denote as $x_k = x(k \tau)$ and $w_k = w(k \tau)$ the sampled states and disturbance at time $k \tau$,   
and let $\phi_k(u) := \phi(u,k\tau)$ and $\psi_k(x) := \psi(x,k\tau)$ for notational brevity. 
We propose the following  projected-gradient-type
controller to generate inputs $\{u_k\}$ at each time $k \tau$, $k \in \mathbb{Z}_{+}$:  
\begin{align}
\label{eq:ideal-controller}
u_{k}  = \proj\left\{u_{k-1} -\eta \Psi_k(u_{k-1}, x_{k}) \right\},
\end{align}
where $\eta > 0$ is a tunable  parameter (also known as step size in the gradient descent literature), and   
$$
\Psi_k(u, x) := \nabla \phi_k(u) 
+ H(u)^\top \nabla \psi_k(x),
$$
where $H(u)$ is the Jacobian of $h_u$ evaluated at $u$.

The controller~\eqref{eq:ideal-controller} is of the form of a projected 
gradient-type algorithm; here, we have modified it by including the gradient evaluated at the instantaneous 
system state $x_k$ to  circumvent the need to measure the exogenous input $w_k$. We also note that the map $\Psi_k$ is applied to the current state $x_k$, and the previous control input $u_{k-1}$, which is applied to the system over the interval $[(k-1) \tau, k \tau)$. Critically, the controller relies on
the knowledge of the system state $x(t)$ at time $\tau k$, which cannot be observed directly.   
To address this, we consider training a neural network to obtain an estimate $\hat p$ of the perception map $p$ (as in Assumption~\ref{as:observable-Map}) which gives estimates of state $x(t)$ from the perceptual data $\zeta(t)$.

Towards this, we consider a set of training points $\{x^{(i)}, \zeta^{(i)} = q(x^{(i)})\}_{i = 1}^N$ to guarantee that the network training is well-posed as stated next. 

\vspace{.1cm}

\begin{assumption}\label{as:trainingSet}
The training points $\{ x^{(i)} \}_{i=1}^N$ are drawn from the compact set $\mathcal{X}_{\text{tr}} := \mathcal{X}_{eq} + \mathcal{B}_n[r_{\text{tr}}]$, $r_0 \leq r_{\text{tr}} \leq r$.  \hfill \QEDB
\end{assumption}

\vspace{.1cm}

Hereafter, let $\mathcal{Q} := q(\mathcal{X}_{\text{tr}})$ denote the perception set associated with the training set, which is a compact set. Assumption~\ref{as:trainingSet} allows us to leverage existing results on the bounds on the approximation error $\sup_{\zeta \in \mathcal{Q}} \|\hat{p}(\zeta) - p(\zeta)\|$ of feedforward neural networks and residual neural networks over the compact set $\mathcal{Q}$; see~\cite{hornik1989multilayer} 
and~\cite{Marchi2022,tabuada2020universal}.

\begin{figure}[h]
    \centering
    \includegraphics[width=\columnwidth]{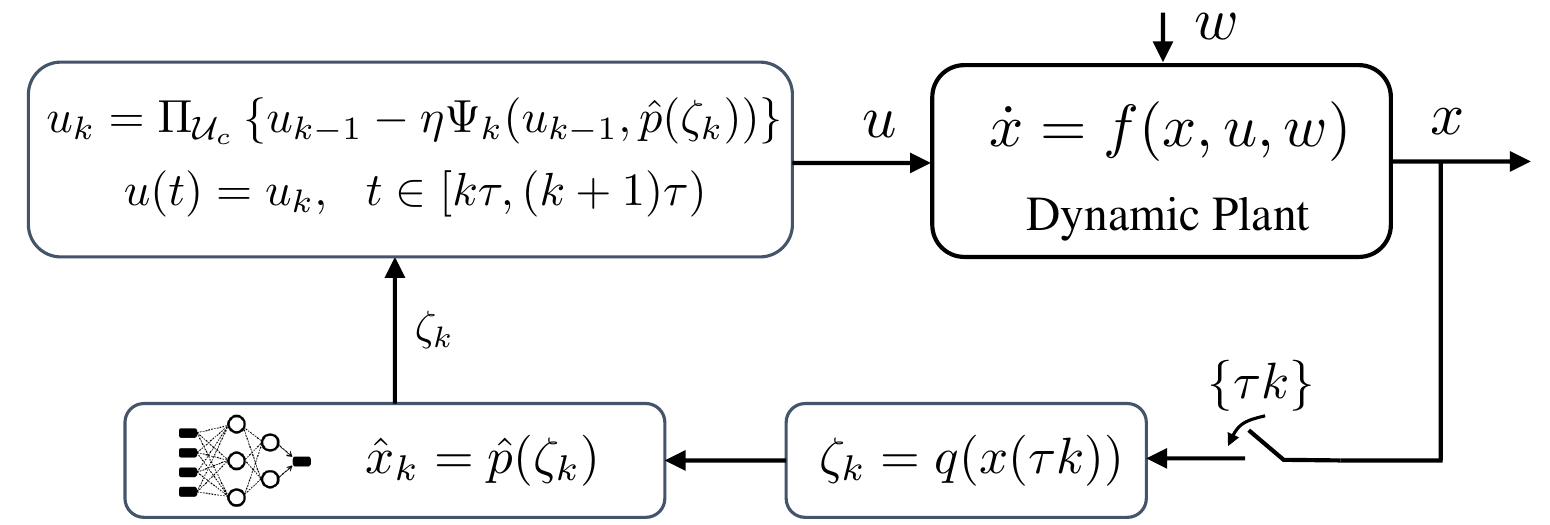}
    \vspace{-.3cm}
    \caption{Block diagram of the proposed perception-based feedback controller in closed loop with the plant. }
    \label{fig:closedLoop}
\end{figure}

With an estimate $\hat p$ of the perception map $p$ obtained via a neural network, the proposed perception-based controller is shown in Figure~\ref{fig:closedLoop} and is tabulated in Algorithm~\ref{alg:perception_cost}. 

%
\begin{algorithm}
\caption{Regulation with NN State Perception}
\label{alg:perception_cost}

\# \textbf{Training}

\quad Given: training set $\{(x^{(i)}, \zeta^{(i)})\}_{i = 1}^N$

\quad Obtain: $\hat p \gets \operatorname{NN-learning}(\{(x^{(i)}, \zeta^{(i))}\}_{i = 1}^N)$

\vspace{.2cm}

\# \textbf{Gradient-based Sampled-Data Feedback Control}

\quad Given: set $\mathcal{U}_c$, funct.s $\phi(\cdot, t), \psi(\cdot, t), H(u)$, $\hat p$, gain~$\eta$

\quad Initial conditions: $x(0) \in \mathcal{X}_0$, $u(0) \in \mathcal{U}_c$

\quad For $t \geq 0$, $k \in \mathbb{Z}_{+}$:
\begin{subequations}
\label{eq:closedloop_perception}
\begin{align}
 \dot x(t) & =  f(x(t), u(t), w(t))  \\
      \zeta_k & = q(x(k \tau)) \\
      u_{k}  & = \proj\left\{u_{k-1} -\eta \Psi_k(u_{k-1}, \hat{p}(\zeta_k)) \right\} \\
      u(t) & = u_{k} , \,\,\,\, t \in [k\tau, (k+1) \tau ) 
\end{align}
\end{subequations}
\end{algorithm}

In the training phase, the operation 
$\operatorname{NN-learning}(\cdot)$ refers to a generic training procedure for 
the neural network via empirical risk minimization, 
%
which results in the approximate map $\hat p (\cdot)$.
In the proposed controller, $\hat p(\cdot)$ is utilized to obtain estimates of the state of the dynamical system $\hat x_k = \hat p(\zeta_k)$, which is subsequently utilized to compute the gradient map $\Psi_k(u, \hat{p}(\zeta))$. 
Note that, as in sampled-data systems, the input $u(t)$ is computed based on the control iterates $\{u_k\}$ as the piece-wise constant signal $u(t)  = u_{k}, t \in [k\tau, (k+1) \tau )$, $k \in \mathbb{Z}_{+}$.

\section{Stability and Tracking Analysis}
\label{sec:stability}

To analyze the performance of the closed-loop system~\eqref{eq:closedloop_perception}, recall that $(u^*_k, x^*_k)$ is the sequence of optimizers of the time-varying problem~\eqref{opt:objectiveproblem} at the times in $\mathbb{S}$. As proposed in~\cite{SampledDataOnlineFeedbackEquilibriumSeeking_2021}, we consider a discrete-time counterpart of~\eqref{eq:closedloop_perception}; sampling~\eqref{eq:closedloop_perception} at times in $\mathbb{S}$ yields:   
\begin{subequations}
\label{eq:closedloop_sampled}
\begin{align}
 x_{k+1} & =  F(x_k, u_k, w_k), \\
      \zeta_k & = q(x_k), \\
      u_{k}  & = \proj\left\{u_{k-1} -\eta \Psi_k(u_{k-1}, \hat{p}(\zeta_k))\right\},
\end{align}
\end{subequations}
where 
$F(X_0, v, w)$ denotes the solution of the initial value problem $\dot X(t) = f(X(t), v, w(t))$, with $X(t_0) = X_0 \in \mathcal{X}_r$ at time $t = t_0 + \tau$, where $\tau$ is the sampling period. The tracking results for~\eqref{eq:closedloop_sampled} will then translate into transient bounds for the sampled-data system~\eqref{eq:closedloop_perception} by using \cite[Theorem 5]{nevsic1999formulas}.

Let $z_k := (x_k - x^*_k, u_k - u^*_k)$ and define the matrix $M_1 \in \R^{2 \times 2}$ as,
\begin{align}\label{eq:matrixM}
    M_1 := \begin{bmatrix}
    c_P & \eta \ell_x \ell_{h_u} c_{w} / \sqrt{d_1} \\
    c_{w} \ell_{h_x} \sqrt{d_1} (1 + c_P) & c_{w}\left( 1 + c_{w} \eta \ell_x \ell_{h_x}^2 \right)
    \end{bmatrix},
\end{align}
where 
$c_{w} := e^{-d_3 \tau /2}\sqrt{d_2/d_1}$, $d_1 > 0$ is given in Assumption \ref{as:class-K-straight-assumption}, and $c_P := \sqrt{1 - \eta(2 \mu - \eta \ell^2)}$; we further define: 
\begin{align}
    M_2 &:= \begin{bmatrix}
    1 & \eta \ell_x \ell_{h_u} \\
    c_w \ell_{h_u} \sqrt{d_1} (c_P + 1) & c_w \sqrt{d_1} \ell_{h_u} \eta \ell_x \ell_{h_u}
    \end{bmatrix}, \\
    M_3 &:= \begin{bmatrix}
    c_w \eta \ell_x \ell_{h_u}^2 (\sqrt{\tau} + 1) \\
    \frac{\eta \ell_x \ell_{h_u}}{\sqrt{d_1}}\sqrt{\tau},
\end{bmatrix} \, .
\end{align}
With these definitions, our main result provides transient bounds for the system~\eqref{eq:closedloop_sampled}.

\vspace{.2cm}

\begin{theorem}[\bf Transient bound for~\eqref{eq:closedloop_sampled}]\label{thm:main-result}
Consider the closed-loop system~\eqref{eq:closedloop_sampled}. Let Assumptions \ref{as:steady-state-map}-\ref{as:costFunctionAssumptions} hold, 
and assume that 
\begin{align}
\label{eq:conditions_thm1}
    \tau > \frac{1}{d_3} \log \left(\frac{d_2}{d_1} \right),  \quad 
    \eta \in \left(0, \frac{2 \mu}{\ell^2}\right).
\end{align}
Then, the tracking error satisfies
%
%
\begin{align}
    \|z_k\| & \leq \frac{r_{M_1} m_2}{m_1} c_{M_1}^{k+1} \|z_0\| + b \|M_3\| \sigma_w\left(\sup_{0 \leq s \leq \tau k} \|\dot w(s)\|\right) \nonumber \\
    & ~~~~~~ + b \|M_2\| \left \lvert \left \lvert 
    \begin{bmatrix}
    \sup_{1\leq s \leq k}\|u_{s}^* - u_{s-1}^*\|\\
    \sup_{\zeta \in \mathcal{Q}} \|\hat{p}(\zeta) - p(\zeta)\| + \bar \varepsilon
    \end{bmatrix}\right \lvert \right \lvert, 
    \label{eq:transient_sampled}
\end{align}
for any $u(0) \in \mathcal{U}_c, x(0) \in \X_{eq} + \B_n(r_I)$ for a sufficiently small $0<r_I<r_0$, where $m_1 := \min \left\{1, \sqrt{d_1}\right\}, m_2 := \max \left\{1, \sqrt{d_2}\right\}$, $b := \frac{r_{M_1} c_{M_1}}{m_1(1 + c_{M_1})}$, and the constants $r_{M_1} > 0$ and $c_{M_1} \in [0,1)$ are s.t. $\|M_1^k\| \leq r_{M_1} c_{M_1}^k, $ $\forall k \in \mathbb{Z}_{+}$.  \hfill \QEDB
\end{theorem}

\vspace{.1cm}

Theorem \ref{thm:main-result} guarantees exponential convergence of the sampled trajectory of the tracking error to a neighborhood of zero. The size of the neighborhood depends on: 
$\sup_{\zeta \in \mathcal{Q}} \|\hat{p}(\zeta) - p(\zeta)\| + \bar \varepsilon$, which corresponds to the error associated with the state estimation,
$\sup_{1\leq s \leq k}\|u_{s}^* - u_{s-1}^*\|$, which captures the time-variability of the optimizer, and 
$\sigma_w(\sup_{0 \leq s \leq \tau k} \|\dot w(s)\|)$ 
corresponds to the time-variability of the unknown disturbance.  The proof is provided in the extended version of the paper~\cite{cothren2022perception_Arxiv}.


The transient bound~\eqref{eq:transient_sampled} shows that the \textit{sampled} system~\eqref{eq:closedloop_sampled} is input-to-state stable (ISS), in the sense of~\cite{jiang2001input}, with respect to $\|\dot w\|$, the drift on the optimal solution $\|u_{k}^* - u_{k-1}^*\|$, and the error introduced by the estimated perception map. 
To translate the results of Theorem~\ref{thm:main-result} into transient bounds for the \textit{continuous} system~\eqref{eq:closedloop_perception}, we leverage Theorem~5 of~\cite{nevsic1999formulas}.  In particular, let $z(t) := (x(t) - x^*(t), u_k - u^*(t))$, where $u^*(t)$ is piece-wise constant and such that $u^*(t) = u^*_k$ for $t \in [k\tau, (k+1)\tau)$, and $x^*(t)$ is defined similarly. Then, we obtain the following.

\vspace{.2cm}

\begin{theorem}[\bf Transient bound for~\eqref{eq:closedloop_perception}]\label{thm:main-resultPerception}
Consider the closed-loop system~\eqref{eq:closedloop_perception}. Let Assumptions \ref{as:steady-state-map}-\ref{as:costFunctionAssumptions} hold, and assume that 
$\tau, \eta$ satisfy~\eqref{eq:conditions_thm1}.
Then, there exist $\beta \in \mathcal{KL}$ and $\gamma_w, \gamma_u, \gamma_p \in \mathcal{K}_\infty$ such that 
\begin{align}
    \|z(t)\| & \leq \beta(\|z_0\|,t) + \gamma_w\left( \sup_{0 \leq s \leq t} \|\dot w(s)\|\right) + \gamma_u\left(\Delta_u^* \right) \nonumber \\
    & ~~~~~~ + \gamma_p\left(
    \sup_{\zeta \in \mathcal{Q}} \|\hat{p}(\zeta) - p(\zeta)\| + \bar \varepsilon \right)
    \label{eq:transient}
\end{align}
holds for any $u(0) \in \mathcal{U}_c, x(0) \in \X_{eq} + \B_n(r_I)$ for a sufficiently small $0<r_I<r_0$, given that $\Delta_u^* := \sup_{k\in \mathbb{Z}_{+}} \|u^*_k - u^*_{k-1}\| \leq r_u$, $\sup_{0 \leq s \leq t} \|\dot w(s)\| \leq r_w$, and $\sup_{\zeta \in \mathcal{Q}} \|\hat{p}(\zeta) - p(\zeta)\| + \bar \varepsilon  \leq r_p$ for some finite $r_w, r_u, r_p > 0$.   \hfill \QEDB
\end{theorem}

Mirroring~\eqref{eq:transient_sampled},  the bound~\eqref{eq:transient} shows that the system~\eqref{eq:closedloop_perception} is ISS  with respect to $\|\dot w\|$, the drift on the optimal solution $\Delta_u^*$, and the error introduced by the neural network. 

To provide a connection with the existing literature, we point out that~\eqref{eq:transient} generalizes the following sub-cases: (i)~when the state $x(t)$ can be observed (without errors), then~\eqref{eq:transient} reduces to a bound similar to~\cite{bianchin2021time} (where, however, the controller is a continuous-time gradient flow); (ii) when the state $x(t)$ can be observed and the functions $\phi(u)$ and $\psi(x)$ are \emph{time invariant}, then~\eqref{eq:transient} boils down to the ISS result of~\cite{SampledDataOnlineFeedbackEquilibriumSeeking_2021} (for sampled-data controllers) and~\cite{MC-ED-AB:20} (for continuous-time gradient flows) and~\cite{zheng2019implicit}. 

\vspace{.1cm}

\begin{remark}
When a residual network is utilized to estimate the map $p$, the error $\sup_{\zeta \in \mathcal{Q}} \|\hat{p}(\zeta) - p(\zeta)\| $ on the compact set $\mathcal{Q}$ in~\eqref{eq:transient} can be bounded as shown in~\cite{Marchi2022,tabuada2020universal}, under given conditions on the selection of the training points. The bound in~\cite{Marchi2022,tabuada2020universal} is particularly interesting because it ties the approximation error with the training error and the geometry of the residual network. We do not include a customization of the bound~\eqref{eq:transient} using the results of~\cite{tabuada2020universal} due to space limitations. 
\QEDB
\end{remark}

\vspace{.1cm}


Theorem~\ref{thm:main-resultPerception} follows from the results of Theorem~\ref{thm:main-result} and Theorem~5 of~\cite{nevsic1999formulas} by noticing that~\eqref{eq:closedloop_perception} is uniformly bounded over an interval $\tau$ in the sense of~\cite[Definition~2]{nevsic1999formulas}. This is because $u(t)$ is piece-wise constant and takes values from the compact set $\mathcal{U}_c$, $\mathcal{W}_c$ is compact  (and $\|\dot w\|$ is bounded), and the perception error is bounded; boundedness of $x(t)$ w.r.t. the sampled sequence $\{x_k\}$  follows from~\cite[Lemma~1]{SampledDataOnlineFeedbackEquilibriumSeeking_2021}. The proof is omitted due to space limitations.

\section{Application to  Autonomous Driving  }\label{S:numerical-example}

We utilize our controller in Algorithm~\ref{alg:perception_cost} to control a vehicle to track a set of reference points while avoiding obstacles; the position of the vehicle is accessible only through camera images. The vehicle's movement is modeled by the unicycle dynamics with state $x = (a, b, \theta)$, where $r := (a,b) \in \R^2$ is the position in the 2D plane and $\theta \in (-\pi, \pi]$ is the orientation with respect to the $a$-axis: $\dot a = v \cos(\theta)$,  $\dot b = v \sin (\theta),\,\, \dot \theta = w$, 
where $v, w \in \R$ are controllable inputs. Importantly, we assume that we do not have direct knowledge of the state $x = (a,b,\theta)$, but instead, camera images, represented by the generative map $\zeta = q(x)$, which return the \textit{position} $r = (a,b)$. We consider a lower-level stabilizer to stabilize the unicycle dynamics to satisfy Assumption \ref{as:class-K-straight-assumption}. Following \cite[Lemma 2]{cothren2022online}, let $u = (u_a, u_b)$ denote the control inputs for each direction $a$ and $b$, and consider the change of coordinates to the error variables, $\xi := \|u - x\|$, $\phi := \operatorname{atan}\left(\frac{u_b - b}{u_a - a} \right) - \theta$. The dynamics of these are $\dot \xi = -v \cos (\phi)$ and $\dot \phi = \frac{v}{\xi} \sin (\phi) - w$. 
By setting 
\begin{align}\label{eq:constants-for-uni-dynamics}
    v = \kappa \xi \cos (\phi), \,\, w = \kappa (\cos(\phi) + 1) \sin (\phi) + \kappa \phi,
\end{align}
for some $\kappa > 0$, the unicycle dynamics admit a globally exponentially stable equilibrium point $(\xi, \phi) = (0, 0)$.  By setting the constants $v, w \in \R$ as in \eqref{eq:constants-for-uni-dynamics}, the dynamic plant satisfies Assumption \ref{as:class-K-straight-assumption}. 

We consider a sequence of locations that the vehicle would like to follow, denoted as  $\mathcal{T}_d :=  \{r_{d,k} \in \mathcal{R}, k \in \mathbb{Z}_{+}\}$, where $\mathcal{R} := \{ r = (a,b) : x = (a,b, \theta) \in \mathcal{X} \}$. To avoid obstacles around the vehicle, we consider building at each time $\tau k$, $k \in \mathbb{Z}_{+}$, the free workspace of the vehicle defined as $\mathcal{F}_k(r_k) := \{ r: a_k^{(i)}(r_k)^\top r - b^{(i)}(r_k) \geq 0,  i = 1, \dots, M_k \}$, 
where $M_k$ is the number of obstacles at  time $\tau k$, $r^{(i)}$ is the center position of the $i$th obstacle, $a_k^{(i)}(r_k) = r^{(i)} - r_{k}$,  $ b^{(i)}(r_k)$ is a scalar computed depending on vehicle and obstacles positions, and $r_k \in \mathcal{R}$  is the position at time $\tau k$. 
The free workspace  $\mathcal{F}(r_k)$ describes a local neighborhood of the vehicle that is guaranteed to be free of obstacles.  

To track the desired trajectory $\mathcal{T}_d$ while avoiding obstacles, we utilize the following 
waypoint-tracking formulation with a barrier
function 
\begin{align*}
    \psi_k(r) := \frac{1}{2}\|r - r_{d,k}\|^2 - \lambda_k \sum_{i=1}^M \log (b^{(i)}(r_k) - a^{(i)}(r_k)^\top x),
\end{align*}
where $\lambda_k>0$ is a tuning parameter; in particular, in the simulations we set   $\lambda_k = 1/\exp(0.1k)$. We set $\phi_k(u) = 0$.

\begin{figure}[t]
    \centering
    \includegraphics[width=\columnwidth]{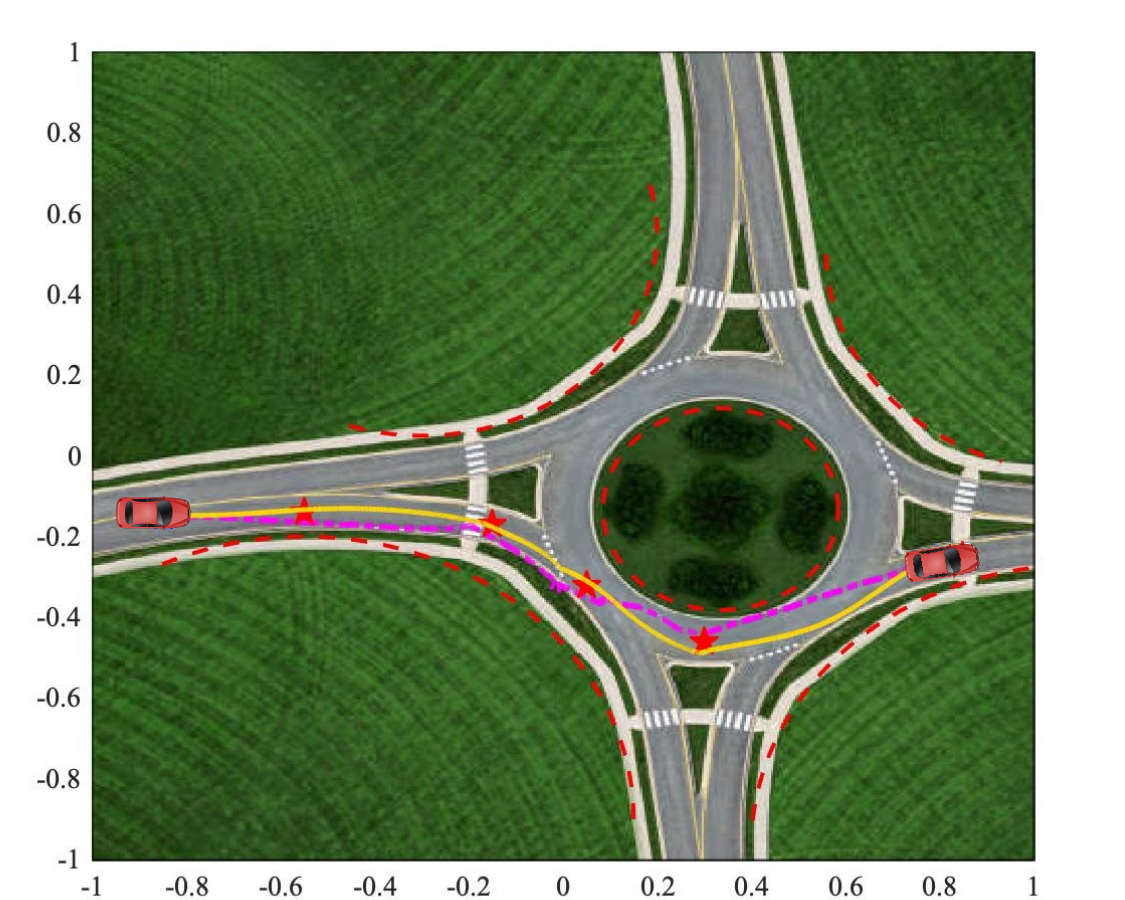}
    \vspace{-.3cm}
    \caption{Sample trajectories of the vehicle tracking the way points represented by red stars. The magenta dashed path is the full execution of Algorithm \ref{alg:perception_cost} with perception, and the yellow path is with perfect state feedback. Red dashed lines identify obstacles used to construct $\mathcal{F}_k.$}
    \label{fig:simulations}
\end{figure}
\begin{figure}[t]
    \centering
    \includegraphics[width=1.0\columnwidth]{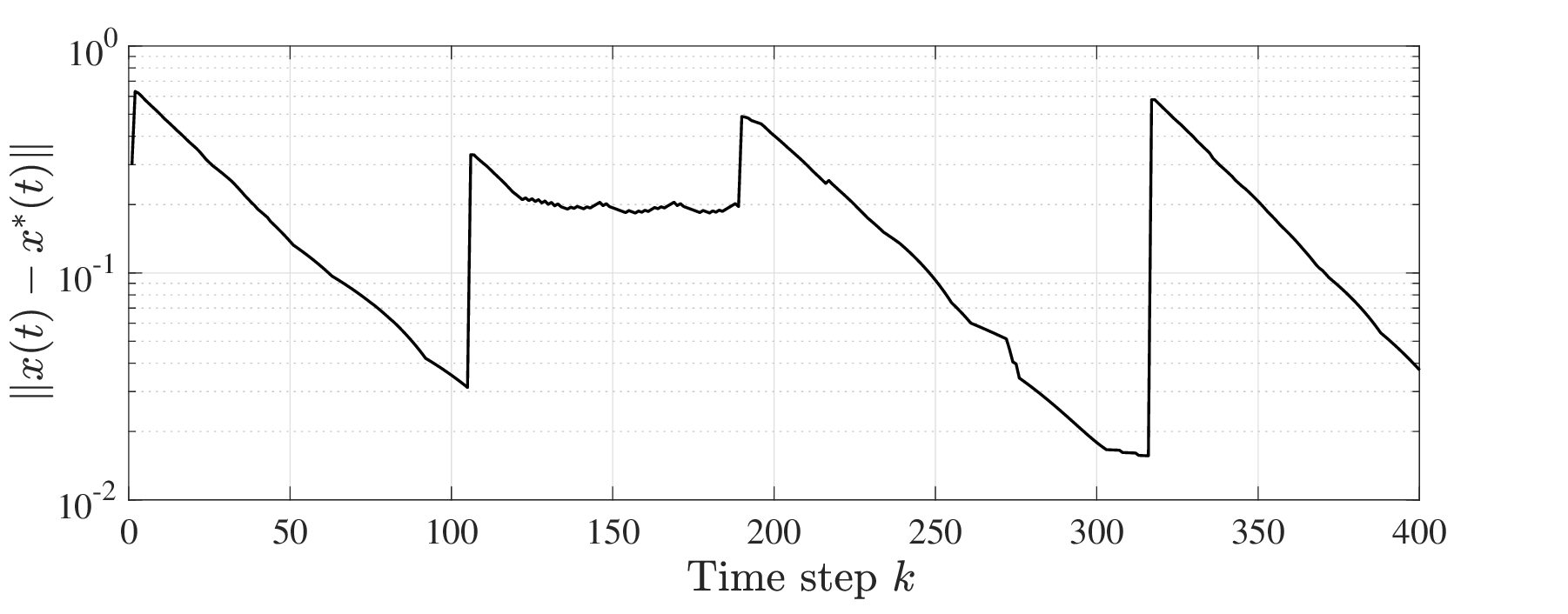}
    \vspace{-.5cm}
    \caption{Error of the position $\|r(t) - r^*(t)\|$. Jumps in the error coincide with variations of the reference $r_{d,k}$. }
    \label{fig:simulations_nn}
\end{figure}

We used a residual neural network to estimate the position of the vehicle from aerial images. specifically, the network returns estimated coordinates in the $(a,b)$. For training, we generate $94,500$ images of size $64 \times 64$ pixels depicting a red bot in the roundabout setting in Figure \ref{fig:simulations}. The images were built using the \textit{MATLAB Image Processing Toolbox} and basic plotting functions therein by setting the background as an aerial view of a roundabout and plotting a red square for the vehicle. We used the \textit{resnet50} structure given in the \textit{MATLAB Deep Learning Toolbox} and tailored the input ($64 \times 64 \times 3$ sized RGB images) and output (total number of labels) sizes to our specific case. For labels, we selected the pixels that corresponded to parts of the image containing the road so that the network only trains on data corresponding to allowable surfaces for the robot, which totaled to $135$ unique labels. Finally, we select five checkpoints along the road for the robot to follow (denoted by the red stars in Figure \ref{fig:simulations}) during the execution of the algorithm corresponding to $\{r_{d,k}\}$. 

Simulation results are given in Fig.~\ref{fig:simulations} for the Algorithm \ref{alg:perception_cost}. The  dashed magenta line corresponds to the trajectory of the unicycle for the neural network-assisted controller, and the yellow line is for the controller with perfect state information. Fig.~\ref{fig:simulations_nn} shows the error $\|r(t) - r^*(t)\|$ between the optimal and actual position of the vehicle. Importantly, the trajectory tracks the time-varying reference points within an error and avoids the obstacles.

\section{Conclusion}\label{S:conclusion}
We proposed an algorithm to regulate dynamical systems towards the solution of a convex optimization problem when we do not have full knowledge of the system states. Specifically, we developed a sampled-data feedback controller that is augmented with a neural network to estimate the state of the system from high-dimensional sensory data. Our results guaranteed exponential convergence of the interconnected system up to an error term dependent on the temporal variability of the problem and the error due to estimating the state from the neural network. 

\bibliography{References/bibliography,References/main_GB,References/biblio_perception}


\appendix

\section{Proofs}

\emph{Proof of Theorem~\ref{thm:main-result}}.

The proof of the Theorem~\ref{thm:main-result} is divided into 
two main parts. In the first part, we construct intermediate 
bounds to be used in the second part of the proof. 

(\textit{Part 1.a: Lyapunov Function}) The first is to leverage the results in Lemma~1 of~\cite{SampledDataOnlineFeedbackEquilibriumSeeking_2021}. By Assumption \ref{as:class-K-straight-assumption}, we have that for a fixed $u \in \U$,
\begin{align}\label{eq:lyap-Florian-As2-bound}
    \dot V(t) \leq - d_3 V(t) + \sigma_w(\|\dot w\|).
\end{align}
Given the initial condition $V(t_0) = V_0$, \eqref{eq:lyap-Florian-As2-bound} implies that:
\begin{align}
    \begin{split}
        V(x(t)&,u,w_t) \leq V_0 e^{-d_3 (t-t_0)} + \int_{t_0}^t \sigma_w(\|\dot w(s)\|) ds
    \end{split}\\
    &\leq V_0 e^{-d_3 (t-t_0)} + \sigma_w(\bar w),
\end{align}
where $\bar w := \sup_{s \in [t_0,t]} \|\dot w(s)\|$. 
Define $V_{k} := V(x_{k}, u_{k}, w_{k})$ and $W_k := \sqrt{V_k}$. Then, from Lemma~1 of~\cite{SampledDataOnlineFeedbackEquilibriumSeeking_2021}, it follows that  
\begin{align*}
a    W_{k+1} \leq  c_w W_k + c_w \ell_{h_u} \sqrt{d_1} \|u_{k+1} - u_k\| + \sqrt{\tau} \sigma_w'(\bar w),
\end{align*}
where $c_w := e^{-d_3\tau /2} \sqrt{d_2/d_1}$, $\ell_{h_u}$ is the Lipschitz constant of the steady state map w.r.t. $u$, and $\sigma_w' := \sqrt{\sigma_w} \in \mathcal{K}$. 

(\textit{Part 1.b: Bound for $\|u_{k+1} - u_k\|$}) To simplify notation, rewrite the controller as $u_{k} := T_k(u_{k-1}, \hat x_k)$, 
where  $T_k(u,x) =  \proj\left\{u -\eta \Psi_k(u, x) \right\}$. Moreover, let $e_{x,k}$ denote the error in the gradient introduced by the perception map. Using this notation, calculate:
%
%
\begin{align*}
    & \|u_{k+1} - u_k\| = \| T_{k+1}(u_k, \hat x_{k+1}) - u_k\|\\
    &= \| T_{k+1}(u_k, \hat x_{k+1}) - T_{k+1} (u_{k+1}^*, h(u_{k+1}^*, w_{k+1})) \\
    & ~~~ + u_{k+1}^* - u_k^* + u_k^* - u_k\|\\
    &= \| T_{k+1}(u_k, \hat x_{k+1}) - T_{k+1} (u_{k}, h(u_{k}, w_{k+1})) \\
    & ~~~ + T_{k+1} (u_{k}, h(u_{k}, w_{k+1})) - T_{k+1} (u_{k+1}^*, h(u_{k+1}^*, w_{k+1})) \\
    & ~~~ + u_{k+1}^* - u_k^* + u_k^* - u_k\|\\
    &\leq \| T_{k+1}(u_k, \hat x_{k+1}) - T_{k+1} (u_{k}, h(u_{k}, w_{k+1}))\| \\
    & ~~~ + \|T_{k+1} (u_{k}, h(u_{k}, w_{k+1})) - T_{k+1} (u_{k+1}^*, h(u_{k+1}^*, w_{k+1}))\| \\
    & ~~~ + \|u_{k+1}^* - u_k^*\| + \|u_k^* - u_k\|\\
    &\leq \eta \ell_x \ell_{h_u} \|x_{k+1} + e_{x,k+1} - h(u_k, w_{k+1})\| \\
    & ~~~ + \|T_{k+1} (u_{k}, h(u_{k}, w_{k+1})) - T_{k+1} (u_{k+1}^*, h(u_{k+1}^*, w_{k+1}))\| \\
    & ~~~ + \|u_{k+1}^* - u_k^*\| +\| u_k^* - u_k\| \\
    &\leq \eta \ell_x \ell_{h_u} \|x_{k+1} - h(u_k, w_{k+1})\| + \eta \ell_x \ell_{h_u} \| e_{x,k+1}\| \\
    & ~~~ + \|T_{k+1} (u_{k}, h(u_{k}, w_{k+1})) - T_{k+1} (u_{k+1}^*, h(u_{k+1}^*, w_{k+1}))\| \\
    & ~~~ + \|u_{k+1}^* - u_k^*\| +\| u_k^* - u_k\|.
\end{align*}
Consider the first term, $\|x_{k+1} - h(u_k, w_{k+1})\|$. We further bound this by applying Assumption \ref{as:class-K-straight-assumption} and by using the fact that $W := \sqrt{V}$:
\begin{align*}
    \|x_{k+1} - h(u_k, w_{k+1})\| &\leq \frac{1}{\sqrt{d_1}} W(x_{k+1}, u_k, w_{k+1})\\
    & \hspace{-.5cm} \leq \frac{1}{\sqrt{d_1}}\left(c_w W(x_k, u_k, w_{k}) + \sqrt{\tau} \sigma_w'(\bar w)\right).
\end{align*}
To bound the second line of the final inequality,
recall that 
\begin{align*} 
\|T_{k+1} &(u, h(u, w_{k+1})) - T_{k+1} (y, h(y, w_{k+1}))\|\leq c_P \|u - y\|
\end{align*} 
holds for any $u,y \in \mathcal{U}$ due to Assumption~3. Thus, 
\begin{align*} 
\|T_{k+1} &(u_{k}, h(u_{k}, w_{k+1})) - T_{k+1} (u_{k+1}^*, h(u_{k+1}^*, w_{k+1}))\|\\
&\leq c_P \|u_{k} - u_{k+1}^*\|\\
&\leq c_P\left(\|u_{k+1}^* - u_{k}^*\| + \|u_{k}^* - u_{k}\| \right).
\end{align*} 

In total, 
\begin{align}
    \begin{split}
        \|u_{k+1} &- u_k\| \leq \frac{\eta \ell_x \ell_{h_u}}{\sqrt{d_1}}\left(c_w W(x_k, u_k, w_{k}) + \sqrt{\tau} \sigma_w'(\bar w)\right) \\
        & ~~~ + \eta \ell_x \ell_{h_u} \| e_{x,k+1}\| \\
    & ~~~ + (c_P + 1)\left(\|u_{k+1}^* - u_{k}^*\| + \|u_{k}^* - u_{k}\| \right).
    \end{split}
\end{align}

(\textit{Part 1.c: Bound for $\|u_{k+1} - u_{k+1}^*\|$}) We obtain the following bound by using similar steps as in the earlier computation to bound $\|u_{k+1} - u_k\|.$ Calculate,
\begin{align*}
    \|&u_{k+1} - u_{k+1}^*\| = \| T_{k+1}(u_k, \hat x_{k+1}) - u_{k+1}^*\|\\
    &= \| T_{k+1}(u_k, \hat x_{k+1}) - T_{k+1}(u_k, h(u_k, w_{k+1})) \\
    & ~~ + T_{k+1}(u_k, h(u_k, w_{k+1})) - u_{k+1}^*\|\\
    &\leq \| T_{k+1}(u_k, \hat x_{k+1}) - T_{k+1}(u_k, h(u_k, w_{k+1}))\| \\
    & ~~ + \|T_{k+1}(u_k, h(u_k, w_{k+1})) - T_{k+1}(u_{k+1}^*, h(u_{k+1}^*, w_{k+1}))\|\\
    &\leq \frac{\eta \ell_x \ell_{h_u}}{\sqrt{d_1}}\left(c_w W(x_k, u_k, w_{k}) + \sqrt{\tau} \sigma_w'(\bar w)\right) \\
        & ~~~ + \eta \ell_x \ell_{h_u} \| e_{x,k+1}\| + c_P\left(\|u_{k+1}^* - u_{k}^*\| + \|u_{k}^* - u_{k}\| \right).
\end{align*}

(\textit{Part 1.d: Overall bounds})  Putting together the bounds above, we have: 
\begin{subequations}\label{eq:summary}
\begin{align}
    \begin{split}
        W_{k+1} &\leq c_w \left( 1 + c_w \eta \ell_x \ell_{h_u}^2 \right) W_k + \sqrt{\tau}( 1 + c_w \eta \ell_x \ell_{h_u}^2 ) \sigma_w'(\bar w)  \\
        & ~~ + c_w \ell_{h_u} \sqrt{d_1} (c_P + 1) \|u_k - u_k^*\| \\
        & ~~ + c_w \eta \sqrt{d_1} \ell_x \ell_{h_u}  \|e_{x,k+1}\| \\
        & ~~ + c_w \ell_{h_u} \sqrt{d_1}(c_P + 1) \|u_{k+1}^* - u_k^*\|.
    \end{split}\\
    \begin{split}
        \|u_{k+1} &- u_{k+1}^*\| \leq \frac{\eta \ell_x \ell_{h_u}}{\sqrt{d_1}}\left(c_w W(x_k, u_k, w_{k}) + \sqrt{\tau} \sigma_w'(\bar w)\right) \\
        & ~~~ + \eta \ell_x \ell_{h_u} \| e_{x,k+1}\| \\
    & ~~~ + c_P (\|u_{k+1}^* - u_k^*\| + \|u_k^* - u_k\|).
    \end{split}
\end{align}
\end{subequations}

Based on these intermediate results, conditions for ISS are derived next. 

(\textit{Part 2: Sufficient conditions for stability}) Define:
\begin{align*}
    \omega_k := \begin{bmatrix}
    \|u_k - u_k^*\| \\
    W_k
\end{bmatrix}, \nu_k := \begin{bmatrix}
    \|u_{k+1}^* - u_k^*\|\\
    \|e_{x,k+1}\|
\end{bmatrix}, \sigma_k := 
\begin{bmatrix}
    0\\
    \sigma_w'(\bar w)
\end{bmatrix}.
\end{align*}
Then, rewrite \eqref{eq:summary} as, 
\begin{align}\label{eq:matrix-summary}
    \omega_{k+1} \leq M_1 \omega_k + M_2 \nu_k + M_3 \sigma_k.
\end{align}
By applying \eqref{eq:matrix-summary} to itself $k+1$ times, we obtain: 
\begin{align}\label{eq:matrix-summary-k+1-times}
    \omega_{k+1} & \leq (M_1)^{k+1}\omega_0 \\
    & ~~~ +  \sum_{s=0}^{k+1} (M_1)^{k+1-s} M_2 \nu_s + \sum_{s=0}^{k+1} (M_1)^{k+1-s} M_3 \sigma_s.
\end{align}
Next, recall from \cite[Ex. 3.4]{jiang2001input} that for a Schur matrix $M_1$, there exists constants $r_{M_1} > 0$ and $c_{M_1} \in [0,1)$ such that $\|(M_1)^k\| \leq r_{M_1} c_{M_1}^k$. By direct computation of the eigenvalues of $M_1$, we may enforce $M_1$ to be Schur when $\tau >0$ satisfies: 
\begin{align}
    \tau > \frac{1}{d_3}\log \left(\frac{d_2}{d_1} \right).
\end{align}
This condition critically relies on the the fact that $\eta > 0$ is chosen so that $c_P \in (0,1)$. Then, taking the norm on both sides (since the quantities are non-negative), and using the triangle inequality, one gets: 
\begin{align}\label{eq:matrix-summary-normed}
\begin{split}
    \| \omega_{k+1} \| &\leq r_{M_1}(c_{M_1})^{k+1}\|\omega_0\| \\
    & +  r_{M_1} \sum_{s=0}^{k+1} (c_M)^{k+1-s}\|M_2\|\|\nu_s\| \\
    & +  r_{M_1} \sum_{s=0}^{k+1} (c_M)^{k+1-s}\|M_3\|\|\sigma_s\|.
    \end{split}
\end{align}
Define the following: 
\begin{subequations}
\begin{align}
    \bar \nu &:= \sup_{0 \leq s \leq k} \|\nu_s\|,\,\, \bar M_2 := \|M_2\|,\\
    \bar \sigma &:= \sup_{0 \leq s \leq k} \|\sigma_s\|,\,\, \bar M_3 := \|M_3\|.
\end{align}
\end{subequations}
Then, \eqref{eq:matrix-summary-normed} becomes, 
\begin{align}\label{eq:MIN_3}
\begin{split}
   \| \omega_{k+1} \| &\leq r_{M_1}(c_{M_1})^{k+1}\|\omega_0\| +  r_{M_1} \left(\sum_{s=1}^{k+1} (c_{M_1})^{s}\right) \bar M_2 \|\bar \nu\| \\
   & +  r_{M_1} \left(\sum_{s=1}^{k+1} (c_{M_1})^{s}\right) \bar M_3 \|\bar \sigma\|.
\end{split}
\end{align}
Equivalently, 
\begin{align}\label{eq:MIN_4}
\begin{split}
   \| \omega_{k+1} \| &\leq r_{M_1}(c_{M_1})^{k+1}\|\omega_0\| \\
   & +  r_{M_1} c_{M_1} \left(\sum_{s=1}^{k} (c_{M_1})^{s}\right)\bar M_2\|\bar \nu\| \\
   & +  r_{M_1} c_{M_1} \left(\sum_{s=1}^{k} (c_{M_1})^{s}\right)\bar M_3\|\bar \sigma\|.
   \end{split}
\end{align}
Then, we may apply the geometric series on the second and third terms to obtain: 
\begin{align}\label{eq:MIN_5}
\begin{split}
   \| \omega_{k+1} \| &\leq r_{M_1}(c_{M_1})^{k+1}\|\omega_0\| \\
   & +  r_{M_1} \frac{c_{M_1}}{1 + c_{M_1}} \left( \bar M_2\|\bar \nu\| + \bar M_3 \|\bar \sigma\| \right).
   \end{split}
\end{align}
Then, finally apply the Lyaunov quadratic bound to obtain the final form.
By Assumption \ref{as:class-K-straight-assumption}, we may write: 
\begin{align}\label{eq:m1m2Bounds}
    m_1 \left \lvert \left \lvert \begin{bmatrix}
    x_{k}  - x_{k}^* \\
    u_k - u_{k}^* 
    \end{bmatrix}\right \lvert \right \lvert \leq \|\omega_k\| \leq m_2 \left \lvert \left \lvert \begin{bmatrix}
    x_{k}  - x_{k}^* \\
    u_k - u_{k}^* 
    \end{bmatrix}\right \lvert \right \lvert,
\end{align}
where 
\begin{align}
    m_1 := \min \left\{1, \sqrt{d_1}\right\}, \,\, m_2 := \max \left\{1, \sqrt{d_2}\right\}.
\end{align}
Substituting \eqref{eq:m1m2Bounds} into \eqref{eq:MIN_5}, we obtain: 
\begin{align}
\begin{split}
    &\left \lvert \left \lvert \begin{bmatrix}
    x_{k}  - x_{k}^* \\
    u_k - u_{k}^* 
    \end{bmatrix}\right \lvert \right \lvert \leq \frac{r_{M_1} m_2}{m_1} (c_{M_1})^{k+1}\left \lvert \left \lvert \begin{bmatrix}
    x_0 - x_{0}^*\\
    u_0 - u_{0}^*
    \end{bmatrix}\right \lvert \right \lvert \\
    & ~~ + \frac{r_{M_1} c_{M_1} \bar M_2}{m_1(1 + c_{M_1})} \left \lvert \left \lvert \begin{bmatrix}
    \sup_{0\leq s \leq k}\|u_{s}^* - u_{s-1}^*\|\\
    \sup_{0\leq s \leq k}\|e_{x,s}\|
    \end{bmatrix}\right \lvert \right \lvert \\
    & ~~ + \frac{r_{M_1} c_{M_1} \bar M_3}{m_1(1 + c_{M_1})} \left \lvert \left \lvert \begin{bmatrix}
    0\\
    \sigma_w(\bar w)
    \end{bmatrix}\right \lvert \right \lvert .
\end{split}
\end{align}
The bound then follows by noticing that:
\begin{align*}
    \|e_{x,k}\| &= \|\hat x_k - x_k\|\\
    &= \|\hat p(\zeta_k) - x_k\|\\
    &= \|\hat p(\zeta_k) - p(\zeta_k)\| + \| p(q(x_k)) - x_k\|
\end{align*}
by using the fact that $\| p(q(x_k)) - x_k\|\leq \varepsilon$, and by bounding the first term with $\sup_{\zeta \in \mathcal{Q}}\|\hat p(\zeta) - p(\zeta)\|$.

\end{document}